\documentclass[twocolumn,aps,prl,groupedaddress,amssymb,showpacs]{revtex4}
\usepackage{graphicx}
\usepackage{dcolumn}
\usepackage{amsmath}
\begin{document}

\title{  Anomalous low temperature ambipolar diffusion and Einstein relation}
\author{A. L. Efros}
\email{efros@physics.utah.edu} \affiliation{Department of Physics,
University of Utah, Salt Lake City UT, 84112 USA}


\begin{abstract}
Regular Einstein relation, connecting the coefficient of ambipolar
diffusion and the Dember field with mobilities, is generalized for
the case of interacting electron-hole plasma. The calculations are
presented for a  non-degenerate plasma injected by light in
semiconductors of silicon and germanium type. The Debye-Huckel
correlation and the Wigner-Seitz exchange terms are considered.
The corrections to the mobilities of  carriers due to difference
between average and acting electric fields within the
electron-hole plasma is  taken into account. The deviation of the
generalized relation  from the regular Einstein relation is
pronounced at low temperatures and can explain anomaly of the
coefficient of ambipolar diffusion, recently discovered
experimentally.
\end{abstract}
 \maketitle

\subsection{ Introduction}
 The Einstein relation for
electrons\cite{LL} $n U d\mu/dn=e D$ connects mobility $U$ with
diffusion coefficient $D$ by the thermodynamic function $d\mu/dn$
and electron charge $e$. Here $\mu$ is chemical potential, $n$ is
electron density. For the case of the Boltzmann gas of
non-interacting electrons, where $d\mu/dn=k T/n$, one gets a
regular form of the Einstein relation, presented by Einstein
\cite{ALB} and von Smoluchowsky \cite{SM} for the Brownian
particles.  Here
  $k T$ is the temperature in energy units.
However, the Einstein relation can be used in a general case of
interacting particles\cite{ef08}.

  The focus of this  paper is   the ambipolar diffusivity under
 condition that interaction between electrons and holes is substantial.
 I have been initiated by the paper
 of Hui Zhao\cite{zhai} who has  measured the coefficient of the   ambipolar diffusivity (CAD) by
 optical method in silicon-on-insulator (SOI) structure with 750 nm
 silicon layer.  The  doping density is $10^{15}{\rm
cm^{-3}}$ while the density $n$ of electron-hole pairs excited by
light is between $(0.5-3)\times 10^{17}{\rm cm^{-3}}$. The
temperature is in the range $90K-400K$.

The regular Einstein relation for the CAD in non-degenerate and
non-interacting electron-hole plasma has a form
\begin{equation}\label{R}
 D_a=\frac{2 k T}{e}\frac{ U_p U_n}{U_p+U_n},
\end{equation}
where  $U_n,U_p$ are mobilities of electrons and holes
respectively. The CAD observed in Ref.\cite{zhai} approximately
follows Eq.(\ref{R}) at $400K> T>300K$. At lower temperatures the
CAD goes down and deviates from the Eq.(\ref{R}) approximately 7
times at $T=90K$. The mobilities were taken from experiments with
a pure bulk silicon.  Author explained this anomalous behavior as
a result of uncontrolled defects that are absent in a bulk
silicon.

I propose here alternative and  more universal intrinsic
explanation of the low temperature anomaly. This explanation is
based upon a novel form of Einstein's fundamental relation for the
CAD, that takes into account interaction between excited carriers.

Under conditions of the experiment the  electron-hole plasma can
be considered as a non-degenerate.   On the other hand, the
classical correlation energy per particle $\sim e^2n^{1/3}/\kappa$
is of the order of 200K. Here $\kappa$ is dielectric constant.
Thus, in the region of the observed anomaly the interaction energy
becomes of the order of temperature.

The paper is organized as follows. First  the novel form of the
fundamental Einstein relation connecting the CAD and the Dember
field with mobilities of electrons and holes is derived. The
relation contains the derivatives of the Helmholtz energy density
(HED) of the interacting plasma with respect to particle
densities. The HED is calculated taken into account correlation
and exchange between particles. Then the corrections
 to the mobilities due to deviation of acting
electric field from the applied field are considered. Finally the
theoretical results  are compared with the experimental data.

\subsection{ Einstein relation for ambipolar diffusivity. Thermodynamics approach}

Silicon is an example of  semiconductor with a long recombination
time of interband excitation. Assume that this system is in the
thermodynamic equilibrium with respect to all relevant parameters
except the total numbers of electrons and holes. To derive the
Einstein relation I use here the same method as in
Ref(\cite{ef08}). The difference is that excited electrons and
holes have two independent electrochemical potentials $\Phi^n$ and
$\Phi^p$ respectively. The Helmholtz energy has a form
\begin{eqnarray}\nonumber
  F &=& \int f(n,p)d^3 r+\int e (p({\bf r})-n({\bf r}))\psi d^3 r \\
  &-& \Phi^p\int p({\bf r})d^3 r-\Phi^n\int n({\bf r})d^3 r.
\end{eqnarray}

Here $n$ and $p$ are electron and hole densities.
 Function $f(n,p)$ is the
HED  of the almost neutral and microscopically homogenous
electron-hole plasma, the function $\psi({\bf r})$ is a potential
of a static electric field.

Using conditions $\delta F/\delta p=0$ and $\delta F/\delta n=0$
one gets
\begin{equation}\label{1}
\Phi^p=\frac{\partial f(n,p)}{\partial p} +e\psi
\end{equation}
and
\begin{equation}\label{2}
\Phi^n=\frac{\partial f(n,p)}{\partial n}-e\psi.
\end{equation}
It follows from the general principles of statistical physics that
that in the state of equilibrium both $\Phi^p,\Phi^n$ should be
constant along the system. Exploring Einstein's idea that electric
field is equivalent to a certain density gradient one can write
the fluxes of holes and electrons as
\begin{equation}\label{3}
{\bf q}_p=-\frac{\sigma_p}{e^2}\nabla \Phi^p,{\bf
q}_n=-\frac{\sigma_n}{e^2}\nabla \Phi^n,
\end{equation}
where $\sigma_p$ and $\sigma_n$ are conductivities of holes and
electrons respectively. Then the fluxes  are
\begin{equation}\label{5}
{\bf q}_p=\frac{\sigma_p}{e}{\bf
E}-\frac{\sigma_p}{e^2}\left(\frac{\partial^2 f}{\partial
p^2}\nabla p +\frac{\partial^2 f}{\partial p\partial n}\nabla n
\right)
\end{equation}
and
\begin{equation}\label{6}
{\bf q}_n=-\frac{\sigma_n}{e}{\bf
E}-\frac{\sigma_n}{e^2}\left(\frac{\partial^2 f}{\partial
n^2}\nabla n +\frac{\partial^2 f}{\partial p\partial n}\nabla p
\right).
\end{equation}
Note that separation of the  conductivities of electrons and holes
are possible only if their mutual scattering is small.
 We assume here that the mobilities of the carriers are
controlled by the lattice scattering. But even in this case the
above expressions predict a drag effect due to the interaction
terms in the HED. The flux of holes is proportional to gradient of
electron density and to gradient of hole density. The same is true
for the flux of electrons. For example,
 if electric field is zero and  gradient of electron density is
 zero, there is an electron flux proportional to a gradient of
  hole density.

 The ambipolar diffusion is measured under condition that electrical
circuit is open. Then the total electric current ${\bf j}$ is zero
and ${\bf q}_p={\bf q}_n$. Taking into account the continuity
equation
\begin{equation}\label{7}
e\partial(p-n)/\partial t+{\rm div}{\bf j}=0.
\end{equation}
one finds that  since electron and holes are excited by light in
equal amounts the system is neutral everywhere. Then $n({\bf
r})=p({\bf r})$ and $\nabla n=\nabla p$. A small separation of
charges at the boundaries of the sample appears due to a
difference  of diffusion coefficients of electrons and holes. This
difference is compensated by  an electric field called the Dember
field. Nevertheless, the electron-hole plasma in the bulk of the
sample is neutral.

The expressions for flaxes can be written in a form
\begin{equation}\label{Dp}
{\bf q_p}=\frac{\sigma_p}{e}{\bf E} -D_p\nabla p
\end{equation}
and
\begin{equation}\label{Dn}
{\bf q_n}=-\frac{\sigma_n}{e}{\bf E} -D_n\nabla n .
\end{equation}

It follows from Eqs.(\ref{5},\ref{6},\ref{Dp},\ref{Dn}) that
diffusion coefficients   of electrons and holes $D_n$ and $D_p$
are connected with corresponding mobilities $U_n,U_p$ by the
relations
\begin{equation}\label{10}
D_n=\left(\frac{\partial^2 f}{\partial n^2}+\frac{\partial^2
f}{\partial n
\partial p}\right)\frac{n U_n}{e}
\end{equation}
and
\begin{equation}\label{11}
D_p=\left(\frac{\partial^2 f}{\partial p^2}+\frac{\partial^2
f}{\partial n
\partial p}\right)\frac{n U_p}{e}.
\end{equation}
Eqs.(\ref{10},\ref{11}) give a novel form of the Einstein relation
that is valid in the case of interacting plasma.

 The mixed partial derivatives in these equations describe the drag
effect.

 Using Eqs.(\ref{Dp},\ref{Dn}) and condition ${\bf q}_p={\bf q}_n$ one gets
expression for the Dember field  in terms of $D_n$, $D_p$
\begin{equation}\label{12}
{\bf E}_D=\frac{e(D_p-D_n)}{\sigma_p+\sigma_n} \nabla n.
\end{equation}
This regular form  becomes more complicated if   $D_n$ and $D_p$
 are expressed through mobilities $U_n$ and $U_p$ using Eqs.(\ref{10}, \ref{11}). Then
\begin{eqnarray}\label{13}
 \nonumber
 {\bf E}_D&=& \left[e(U_n+U_p)\right]^{-1}\left(\frac{\partial^2 f}{\partial n^2}U_n -
 \frac{\partial^2 f}{\partial p^2}U_p\right.\\
  &+&\left.\frac{\partial^2f}{\partial n \partial p}(U_n-U_p)\right)\nabla n.
\end{eqnarray}
Due to the interaction of carriers the Dember field is not
necessarily proportional to
 the difference of mobilities $U_n-U_p$.

Substituting Eq.(\ref{12}) into Eqs.(\ref{Dp},\ref{Dn}) one gets
\begin{equation}\label{12}
{\bf q_n=q_p}=-D_a \nabla n,
\end{equation}
where the CAD
\begin{equation}\label{13a}
D_a=\frac{D_n U_p+D_p U_n}{U_p+U_n}.
\end{equation}
Using Eqs.(\ref{10},\ref{11}) we get  the generalized Einstein
relation for the CAD
\begin{equation}\label{14}
D_a=\frac{2 k T}{e}\frac{ U_p U_n}{U_p+U_n}Q(n,T),
\end{equation}
where
\begin{equation}\label{15}
 Q(n,T)=\frac{ n}{2 k T}\left(\frac{\partial^2 f}{\partial
n^2}+\frac{\partial^2 f}{\partial p^2}+2\frac{\partial^2
f}{\partial n
\partial p}\right)
\end{equation}
is a ratio of the coefficients of ambipolar diffusivity calculated
with and without interaction (cp Eq.(\ref{14}) with Eq.(\ref{R})).

It is important to put $n=p$ after calculation of the second
partial derivatives  in Eqs.(\ref{10},
\ref{11},\ref{13},\ref{15}).

\subsection{ HED of semiconductors with band
structure of Si and Ge}

Analytical calculations of the HED of interacting carriers are
possible in the  framework of  perturbation theory only. One
should keep in mind, however, that the region  of applicability of
these calculations does not cover all  temperature range of the
experiment.

The HED can be written in a form $f=f_{id}+ f_i$, where the first
term describes the ideal gas, while the second one takes into
account interaction. The non-interacting carriers are independent
and $f_{id}= f^p_{id}(p)+f^n_{id}(n)$. Since  only the second
derivatives of the HED are necessary, one can write  $f_{id}(p)=p
k T (1+\ln p)$ and  $f_{id}(n)=n k T (1+\ln n)$.

The largest interaction term for the non-degenerate plasma
describes correlation effect. It was calculated by Debye and
Huckel\cite{Debye} in a form
\begin{equation}\label{16}
f_c(n+p)=-(2e^3/3 \kappa^{3/2})\sqrt{\pi/{k T}}(n+p)^{3/2}.
\end{equation}
 This term is independent of
the spectra of electrons and holes. In our approximation this is the
only term that has non-zero $\partial^2 f/(\partial n\partial p)$
and contributes to the drag effect (See Eqs.(\ref{5},\ref{6})).

I also take into account the exchange interaction between
electrons in each ellipsoid, between heavy holes and between light
holes. This interaction term has a higher power of $T$ in the
denominator of the HED than the correlation term. The
thermodynamic potential density $\Omega(\mu, T)$ for this
interaction can be written in a form of the Wigner-Seitz integral
(See\cite{LLS})
\begin{equation}\label{17}
\Omega_{ex}=-\frac{4 \pi e^2 }{\hbar^4 \kappa}\int\int
\frac{n_{p1}n_{p2}d^3p_1d^3p_2}{(\vec{p_1}-\vec{p_2})^2 (2\pi)^6},
\end{equation}
where $n_p$ is the Fermi function that  has the Boltzmann form in
this case. To find the HED one should express chemical potentials
through the density of carriers.

 For a conduction band consisting
of $g$ equivalent ellipsoids of rotation  one gets
\begin{equation}\label{18}
 f_{ex}(n)=-\frac{n^2 e^2\hbar^2 I(a)}{4\kappa\sqrt{\pi} g m_{\perp}k T},
\end{equation}
where  masses $m_{\perp}$ and $m_{\parallel}$,  are perpendicular
and parallel to the rotation axis of an ellipsiod,
$a=m_{\parallel}/m_{\perp}$. At $a>1$
\begin{equation}\label{20}
I(a)=\frac{2\sqrt{\pi}\arctan(\sqrt{a-1}}{\sqrt{a-1}}.
\end{equation}

 For a parabolic valence band with light hole $m_l$ and heavy
hole $m_h$
\begin{equation}\label{21}
f_{ex}(p)=-\frac{\pi e^2 \hbar^2 p^2(m_h^2+m_l^2)}{2 \kappa k
T(m_h^{3/2}+m_l^{3/2})^2}.
\end{equation}

 My final
 result for corrections to a regular Einstein relation reads

\begin{eqnarray}
 \nonumber
  Q(n,T)&=& 1-\frac{e^3\sqrt{\pi n} }{\sqrt{2}\kappa^{3/2}(k T)^{3/2}}- \frac{e^2\hbar^2 n I(a)}
  {4\sqrt{\pi}\kappa m_{\perp}
g (k T)^2}\\
   &-& \frac{\pi e^2 \hbar^2 n (m_h^2+m_l^2)}{2\kappa (k T)^2(m_h^{3/2}
+m_h^{3/2})^2}.
\end{eqnarray}

The Dember field in terms of mobilities has a form
\begin{eqnarray}\label{23}
 \nonumber
  E_D &=& \frac{1}{(eU_n+eU_p)}\left(\left(\frac{k T}{n}+\frac{e^3\sqrt{\pi}}{\kappa^{3/2}\sqrt{2k T n}}\right)(U_p-U_n)\right .\\
      &-& \left .\frac{\pi e^2 \hbar^2 (m_h^2+m_l^2)U_p}{\kappa k T(m_h^{3/2}+m_l^{3/2})^2}+\frac{ e^2\hbar^2 I(a)U_n}{2\kappa\sqrt{\pi} g m_{\perp}
     k T}\right)\nabla n.
\end{eqnarray}

  \subsection{ Mobility of carriers in a plasma}
It is assumed above that mobilities of the carriers are controlled
by the lattice scattering. But even in this case there are
important corrections to these mobilities due to the interaction
of electrons and holes. Each carrier is surrounded by a screening
atmosphere of the opposite sign. This atmosphere is polarized by
an applied electric field. The field of this polarization is
opposite to the applied field so that the effective field acting
on the carrier is less than applied field. It can be interpreted
as a decrease of the mobility. The theory of this effect was
created by Debye, Huckel, and Onsager (See Ref.\cite{LLK}). The
resulting changes of the mobilities are
\begin{equation}\label{24}
S(n,T)=\frac{\Delta U_n}{U_n}=\frac{\Delta U_p}{U_p}=
-\frac{\sqrt{2\pi}e^3 n^{1/2}}{3(1+\sqrt{0.5})\kappa^{3/2}(k
T)^{3/2}}.
\end{equation}
Coming back to the Einstein relation Eq.(\ref{14}) one should note
that if the  mobilities $U_n,U_p$ are measured in the presence of
plasma, the above corrections  are irrelevant because  the
experimental values $U_n,U_p$ contain them. However, if the
mobilities are known from experiments without light excitation, as
in the case of Ref.\cite{zhai}, the Einstein relation
Eq.(\ref{14}) takes a form
\begin{equation}\label{25}
D_a=\frac{2 T}{e}\frac{ U_p U_n}{U_p+U_n}P(n,T),
\end{equation}
where $P(n,T)=Q(n,T)+S(n,T)$.
 One can see that this change increases the numerical factor in the
second term of $Q$, originated from correlation, by 1.39.

\subsection{Discussion and Conclusion}

  \begin{figure}
  \includegraphics[width=7 cm]{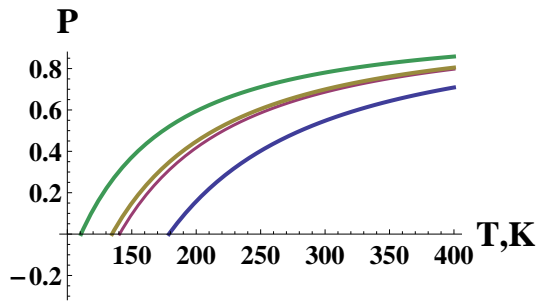}\\
  \caption{(Color online)Function $P(n,T)$ for Si at three different
  values of $n_0$ defined as
  $n=n_0\times 2.3\times 10^{17} \rm{cm}^{-3}$; $n_0= 2$ for lower curve (blue),
  $n_0=1$ for
   two intermediate curves, and $n_0= 0.5$ for upper curve(green).
    The upper intermediate curve  does not take
   into account exchange interaction.}
\end{figure}
\begin{figure}
  \includegraphics[width=7 cm]{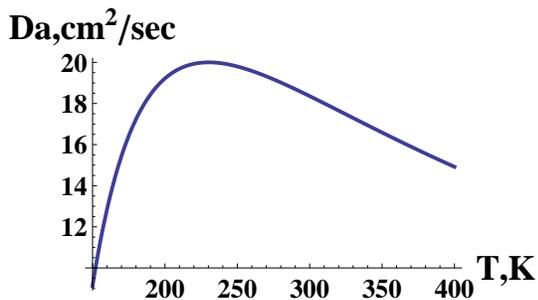}\\
  \caption{(Color online)Temperature dependence of $D_a$ obtained under assumption that $U_p\sim
  T^{-2.2}$ at $n=2.3\times 10^{17} \rm{cm}^{-3}$. The absolute values of the hole  mobility is chosen  such
  that maximum value of  $D_a=20 {\rm cm^2/sec}$. Maximum occurs
  at $T\approx 230K$, which is very close to the experimental
  result.}
\end{figure}
Now I discuss the low temperature anomaly of the CAD  in Si.
Function $P(n,T)$ is shown in Fig.1 in a proper temperature range.
The comparison of the first approximation (correlation) with the
second one (exchange) shows that the perturbation theory looks
reasonable at $T\geq 150$K and at $n_0=1$.

To estimate CAD  as a function of $T$ one should know mobilities
$U_n$ and $U_p$.  The experimental and theoretical data of
Ref.\cite{jac} show that in silicon $U_p/U_n\approx 0.25$, at
$T\approx 200$K. Then $U_pU_n/(U_p+U_n)\approx U_p$.   The hole
mobility in the  pure silicon is due to the phonon scattering and
it depends on temperature as $ T^{-2.2}$ in all temperature range
considered. The deviation from a usual law $T^{-1.5}$ is due to
the warping of the top of the valence band. The mobility of doped
silicon with the hole density $2\times 10^{17} cm^{-3}$ has
$T^{-1.5}$ dependence\cite{jac} at $T>180K$ that may be
interpreted as a phonon scattering but with the warping  smeared
by the doping.  A pure silicon is considered here,  and $T^{-2.2}$
mobility dependence is used.

 The expression $D_a= P(T)U_p T$ with $U_p=R T^{-2.2}$ is used to get
T-dependence of $D_a$ that follows from the above theory. To make
comparison with experimental result easier the factor $R$ is
chosen such that $D_a=20{\rm cm^2/sec}$ in the maximum, similar to
the experimental data of Ref\cite{zhai}. The theoretical result is
shown in Fig. 2.

Since the HED is calculated using  perturbation theory, the
discussion of the low temperature behavior might be doubtful, and
the most important argument is position of the maximum of CAD.
Clearly the way factor $R$ is chosen has no effect on the maximum
position. Theoretical position of maximum is 230K, which is close
to the experimental position that has some uncertainty because of
the large error bars. I think this similarity is a strong argument
in favor of the proposed explanation.

 In the range $T>150K$  the the theoretical  curve is similar
to experimental points of Ref(\cite{zhai}). At lower $T$
theoretical values become negative. This definitely means collapse
of the perturbation theory, because negative CAD leads to an
absolute instability of a neutral plasma\cite{ef08}. It is
intriguing that just near this temperature the character of
experimental data changes: CAD becomes independent of both $T$ and
$n$. This might be a manifestation of a new  phase. The
speculations about this phase are outside the scope of this paper.
I would only mention that amount of excitons at these
temperatures, as given by the Saha equation, is negligible but the
Saha equation is not reliable for the case of  non-ideal plasma.
It is known also that
 the phase of exciton gas-liquid coexistence
corresponds to  lower temperature at these densities\cite{kittel}.

Finally, I proposed an  explanation of the low temperature anomaly
of the CAD based upon the Einstein relation for interacting
carriers. The applicability of the theory is limited at  low
enough temperatures because the calculation of the HED is
perturbational. Using fundamental thermodynamics relations
Eqs.(\ref{12},\ref{14},\ref{15}) one could restore unknown
thermodynamic functions of the non-ideal plasma at low
temperatures by measuring mobilities, CAD, and the Dember field.

I am grateful to M.I.  Dyakonov for an important critical comment
and to Hui Zhao for a valuable discussion.

\end{document}